\documentclass[aps,twocolumn,nopacs,superscriptaddress]{revtex4}

\usepackage{graphicx}
\usepackage{dcolumn}
\usepackage{bm}
\usepackage{hyperref}
\usepackage{amsmath}
\usepackage[usenames]{color}
\newcommand{\beq}{\begin{equation}}
\newcommand{\eeq}{\end{equation}}
\newcommand{\beqa}{\begin{eqnarray}}
\newcommand{\eeqa}{\end{eqnarray}}

\begin{document}

\title{ The Chaos Within Sudoku }

\author{M\'aria Ercsey-Ravasz} \email{ercsey.ravasz@phys.ubbcluj.ro}
\affiliation{Faculty of Physics, Babe\c{s}-Bolyai University, 
Str. Kogalniceanu Nr. 1, RO-400084 Cluj-Napoca, Romania}
\author{Zolt\'an Toroczkai} \email{toro@nd.edu}
\affiliation{Interdisciplinary Center for Network Science and 
Applications (iCeNSA)}  
\affiliation{Departments of Physics,  Computer Science and Engineering, 
University of Notre Dame, Notre Dame, IN, 46556 USA}

\date{\today}

\begin{abstract}
The mathematical structure of  the widely popular Sudoku puzzles is akin to 
typical hard constraint satisfaction problems that lie at the heart of many 
applications, including protein folding and the general problem of finding 
the ground state of a glassy spin system.  Via an exact mapping of Sudoku 
into a deterministic, continuous-time dynamical system,  here we show that 
the difficulty of Sudoku translates into transient chaotic behavior exhibited by 
the dynamical system. In particular, we show that the escape rate  $\kappa$, 
an invariant characteristic of transient chaos, provides a single scalar measure 
of the puzzle's hardness, which correlates well with human difficulty level 
ratings. Accordingly,  $\eta = -\log_{10}{\kappa}$ can be used to define a 
``Richter''-type scale for puzzle hardness, with easy puzzles falling in the range 
$0 < \eta \leq 1$, medium ones within $1 < \eta \leq 2$, hard in $2 < \eta \leq 3$ 
and ultra-hard with $\eta > 3$. To our best knowledge, there are no known 
puzzles with $\eta > 4$. 
\end{abstract}


\maketitle

In Sudoku, considered as one of the world's most popular puzzles \cite{RT11}, 
we have to fill in the cells of a $9\times 9$ grid with integers 1 to 9 
such that in all rows, all columns and in nine $3 \times 3$ blocks  every digit 
appears exactly once, while respecting a set of previously given digits in 
some of the cells (the so-called clues). Sudoku is an exact cover 
type constraint satisfaction problem \cite{GJ90} and it is one of Karp's 21 
NP-complete problems \cite{K72}, when generalized to $N\times N$ grids 
\cite{IEICE_YS03}.  NP-complete problems are ``intractable'' (unless P=NP) 
\cite{GJ90,CommACM_F09} in the sense that all known algorithms that compute 
solutions to them do so in exponential worst-case time (in the number of 
variables $N$); in spite of the fact that if given a candidate 
solution, it takes only polynomial time to check its correctness. 

The intractability of NP-complete problems has important consequences, 
ranging from public-key cryptography to 
statistical mechanics. In the latter case,  for the 
ground-state problem of Ising spin glasses ($\pm1$ spins), one needs to 
find the lowest energy configuration among all the $2^N$ possible 
spin configurations. Additionally, to describe the statistical behavior of such 
Ising spin models, one has to compute the partition function, which is
a sum over all the $2^N$ configurations. Barahona \cite{JPhysA_B82}, then 
Istrail \cite{STOC_I00}
have shown that for non-planar crystalline lattices, the ground-state problem and 
computing the partition function are NP-complete \cite{STOC_I00}. Since there is little hope 
in providing polynomial time algorithms for NP-complete problems, the focus 
shifted towards understanding the nature of the complexity forbidding fast 
solutions to these problems. There has been considerable work in this
direction, especially for the Boolean satisfiability problem $k$-SAT, which 
is NP-complete for $k\geq 3$. Due to completeness, all problems in NP 
(hence Sudoku as well),
can be translated (in polynomial time) and formulated as a $k$-SAT 
problem.    
In $k$-SAT we are given $N$ Boolean variables 
to which we need to assign 0s or 1s (TRUE or FALSE) such that a given 
set of clauses in conjunctive normal form are all satisfied (evaluate to TRUE). 
Just as for the spin glass model, here we also have exponentially many 
($2^N$) configurations or assignments to search.

In the following we treat algorithms  as {\em dynamical systems}. 
An algorithm is a finite set of instructions acting in some state space, 
applied iteratively from an initial state until an end state is reached. 
For example, the simplest algorithm for the Ising model ground state 
problem, or the $3$-SAT problem would be exhaustively testing 
potentially all the $2^N$ configurations, which quickly becomes forbidding 
with increasing $N$. To improve performance, algorithms have become 
more sophisticated by exploiting the {\em structure of the problem}
(of the state space). Accordingly, now $3$-SAT can be solved by a deterministic 
algorithm with an upper bound of $O(1.473^N)$ steps 
\cite{TheorCompSci_BK04}. Here we will only deal with 
deterministic algorithms that is, once an initial state is given, the 
``trajectory'' of the dynamical system is uniquely determined. 
Thus, we expect that the dynamics of those algorithms 
that exploit the structure of hard problems 
will reflect the complexity inherent in the problem itself. Complex behavior by 
{\em deterministic} dynamical 
systems is coined chaos in the literature 
\cite{OttBook, ChaosPhys,TransientChaosBook}, 
and thus the behavior of 
algorithms for hard problems is expected to appear 
highly irregular or chaotic \cite{PNAS_ERT07}.
\begin{figure*}[htbp] \begin{center}
\includegraphics[width=0.9\textwidth]{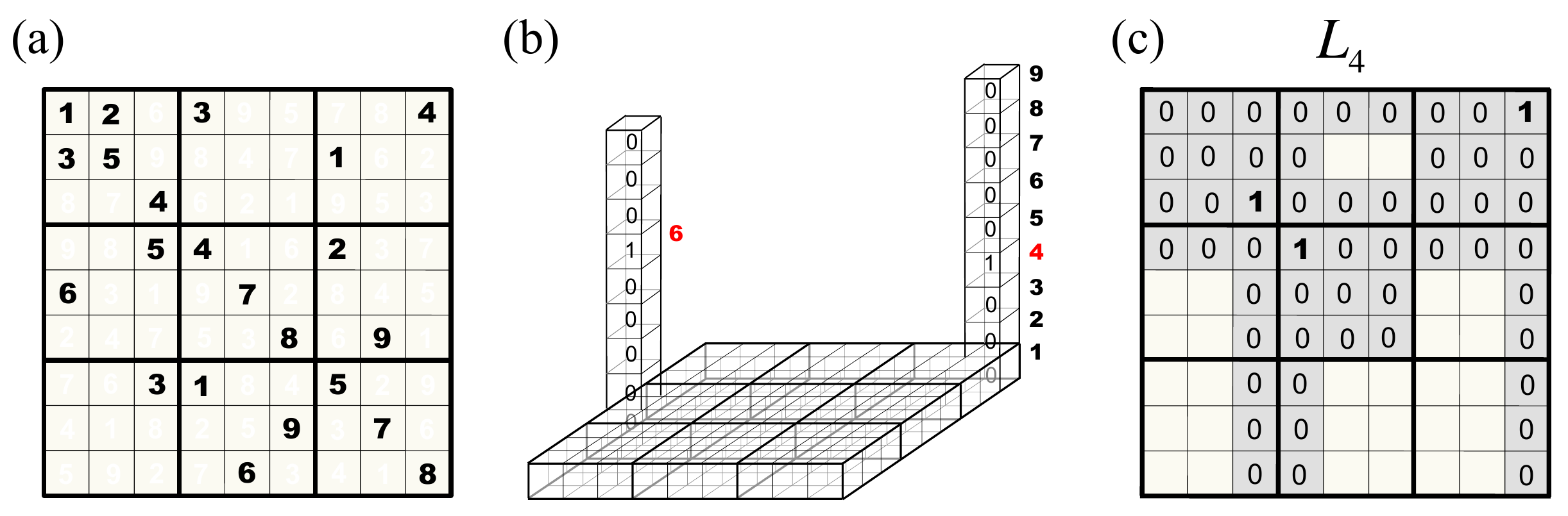}
\caption{{\bf Sudoku and its Boolean representation.} (a) a typical puzzle 
with bold digits as clues (givens). (b) Setup of the Boolean representation 
in a $9\times 9 \times 9$ grid. (c) Layer $L_4$ of the puzzle (the one 
containing the digit $4$) with 1-s in the location of the clues and the 
regions blocked out for digit $4$ by the presence of the clues (shaded area). } \label{fig1} 
\vspace*{-0.5cm} \end{center} 
\end{figure*}

Although the theory of nonlinear dynamical systems and chaos is 
well-established, it has not yet been exploited in the context of optimization 
algorithms. One of the difficulties lies with the fact that most optimization
algorithms are discrete and not easily cast in forms amenable to chaos theory
methods. Recently, however, we have provided 
\cite{NatPhys_ET11} a {\em 
deterministic} continuous-time solver 
 for the Boolean satisfiability problem $k$-SAT
using coupled ordinary differential equations (ODE) with a one-to-one 
correspondence between the $k$-SAT solution clusters and the attractors 
of the corresponding system of ODEs. This continuous-time dynamical system 
(CTDS) is in a form naturally suited for chaos theory methods, and thus it allows
us to study the relationship between optimization hardness and chaotic behavior.
Here we will focus only on solvable (SATisfiable) instances, and thus the observed
chaotic behavior will necessarily be transient \cite{PNAS_KT84,PhysRep_TL08,TransientChaosBook}. 
We need to emphasize, however, that the dynamical properties characterize 
both the problem {\em and the algorithm} itself. For this reason, one 
compares the dynamical properties across problems of varying hardness 
using the same algorithm. Nevertheless, since there are problem instances that
are hard for all known algorithms, the appearance of transient chaos should
be a universal feature of hard problems. It is also important to observe that 
transient chaos is not an $N\to \infty$ asymptotic behavior, but it appears 
for finite $N$, and thus measures of chaos can be used to characterize and 
categorize the hardness of {\em individual instances} of finite problems. To illustrate 
this, here we first map the popular $9\times 9$ (hence finite) version of Sudoku into 
$k$-SAT, then we solve it using our deterministic continuous-time solver 
\cite{NatPhys_ET11}. By analyzing the behavior of the corresponding trajectories 
of the CTDS we show the appearance of transient chaos when increasing the hardness
of the Sudoku problems, and show that the level of hardness (taken from human ratings
of the puzzles) correlates well with a chaotic invariant, namely the {\em  lifetime of chaos} 
$\kappa^{-1}$, where $\kappa$ is called the escape rate \cite{TransientChaosBook}. 
We conclude with a discussion on algorithmic performance, dynamical properties and 
problem complexity. \\

\section*{\large Results}

\subsection*{Sudoku as $k$-SAT}

Because our continuous-time dynamical system \cite{NatPhys_ET11} was designed to 
solve $k$-SAT formulae in conjunctive normal form (CNF), we first briefly describe how 
Sudoku can be interpreted as a $+1$-in-$9$-SAT formula, and then how it is 
transformed into the standard CNF form. Further details are shown in the Methods 
section.

In a Sudoku puzzle we are given a square grid with  $9\times 9 = 81$ cells, each to be filled
with  one of nine symbols (digits) $D_{ij} \in \{1,\ldots,9\}$, $i,j=1,\dots, 9$
(with the upper-left corner of the puzzle corresponding to $i=1$, $j=1$). When 
the puzzle is completed each of the columns, rows and $3\times 3$ sub-grids 
(blocks partitioned by bold lines, Fig. \ref{fig1}a) must contain all the 9 symbols. Equivalently, 
all 9 symbols must appear once and only once in each row, column and $3\times 3$ sub-grid. 

To formulate Sudoku as a constraint satisfaction problem (CSP) using Boolean variables, 
we associate to each symbol (digit) an ordered set of $9$ Boolean variables 
(TRUE=``1'', FALSE=``0''). The digit $D_{ij}$  in cell $(i,j)$ will 
be represented as the ordered set $(x_{ij}^1,\ldots,x_{ij}^9)$ with 
$x_{ij}^{a}\in\{0,1\}$, $a=1,\dots,9$, such that always one and only 
one of them is 1 (TRUE). Thus $D_{ij}=a$ is equivalent to 
writing $x_{ij}^b=\delta_{a,b}$, 
where $\delta_{a,b}$ is the Kronecker delta function. This way we 
have in total $9\times 9\times 9=729$ Boolean variables $x_{ij}^{a}$, 
which we can picture as being placed on a 3D grid  (Fig.~\ref{fig1}b), 
with $a$ corresponding to the grid index along the vertical direction, and hence $a$ 
is the digit that is filling the corresponding $(i,j)$ cell in the original puzzle. 
The corresponding $9\times 9$ 2D layer at height $a$ will be denoted by $L_a$.
For example in the puzzle shown in Fig.~\ref{fig1}a $D_{1,9}=4$. 
In the given vertical column the variable in the $a^{\text{th}}$ 
cell is $x_{1,9}^a=\delta_{a,4}$.
The Sudoku constraints can also be simply encoded using Boolean variables
(see Methods). 
They come from: 1) uniqueness of the symbols in all the $(i,j)$ Sudoku 
cells, 2) a symbol must occur once and only once in each row,
column and in each of the nine $3\times 3$ subgrids, and 3) obeying the clues. 
Constraint type 1) was already expressed above, namely that for every 
cell $(i,j)$, in the set $(x_{ij}^1,\ldots,x_{ij}^9)$ one and only one variable
is TRUE, all others must be FALSE. Type 2) constraints are similar, e.g., 
in row $i$ and layer $a$ the set $(x_{i1}^a,\ldots,x_{i9}^a)$ must contain one and only
one TRUE variable, all others must be false and this must hold for all rows and layers, etc.
Observe that all constraints are in the form of a set of 9 Boolean variables of which
we demand that one and only one of them be TRUE, all others FALSE. When this is
satisfied, we say that the constraint itself (or ``clause'') is satisfied, or TRUE. 
Such CSPs are called $+1$-in-$k$-SAT and they 
are part of  so-called ``locked occupation problems'', which is a class of 
exceptionally hard CSPs \cite{PRL_ZM08,JSMTE_ZM08}. 
Type 3) constraints are generated by the clues (or givens) 
which are symbols already filled in some of the cells 
and their number and positioning determines the difficulty of the puzzle. They
are also set in a way to guarantee a unique solution to the whole puzzle. If
there are given $d$ clues, then this implies setting $d$ Boolean variables
to TRUE, which means eliminating exactly $4d$ constraints of type 1) and 2)
(one vertical or uniqueness constraint, one row, one column and one $3\times 3$
subgrid constraint). Thus, Sudoku is a $+1$-in-9-SAT type CSP with $N$ 
Boolean variables and $324-4d$ constraints. $N$
is a complicated function of the positioning of the clues. 

In order to apply our continuous-time SAT solver we need to bring the $+1$-in-9-SAT 
type CSP above into conjunctive normal form.
In $k$-SAT there are $N$ Boolean variables $x_i=\{0,1\}$ and an instance  
is given as a propositional formula $\mathcal{F}$, which is the conjunction (AND,  
denoted by $\wedge$) of $M$ clauses (constraints) $C_m$:
$\mathcal{F}= C_1 \wedge \dots \wedge C_m \wedge \dots \wedge C_M$\;.
Each clause is the disjunction (OR, denoted  by $\vee $) of  $k$ literals. 
A literal is a variable ($x_i$) or its negation ($\overline x_i $). 
For example a $3$-SAT  constraint 
could  be $C_1= x_1 \vee \overline x_4 \vee x_5$. 
All Boolean propositions $\mathcal{F}$ can be 
formulated in CNF.  

Once the transformation to CNF  is completed we are left with $N$ variables and $M$ 
SAT clauses (see Methods). We will denote the number
of variables appearing in constraint $m$ by $k_m, \, m=1,\dots, M$ (clearly, $1 \leq k_m \leq 9$). 
The parameters $N$, $M$ and $\{k_m\}_{m=1}^{M}$ all depend on the
clues that are difficult to express analytically, but 
easy to determine computationally, as illustrated via examples.

\subsection*{The continuous-time deterministic $k$-SAT solver}

In Ref \cite{NatPhys_ET11} a continuous-time deterministic solver was introduced 
to solve $k$-SAT problems in conjunctive normal form.  The set of clauses 
specifying the constraints are translated into an $M\times N$ matrix: 
$\bm{C} = \{c_{mi}\}$ with $c_{mi}=1$ if the variable $x_i$ is present in 
clause $m$ in direct (non-negated) form, namely
$x_i \in C_m$, $c_{mi}=-1$ if  $\bar{x}_i \in C_m$ and $c_{mi}=0$ if $x_i$ and $\bar{x}_i$
are both absent from $C_m$. To every variable $x_i$ one associates a  continuous spin 
variable $s_i \in [-1,1]$ such that  when $s_i = \pm 1$ then 
$x_i = (1+s_i)/2 \in \{0,1\}$, and to  every clause $C_m$ one associates the function:
\begin{equation}
K_m(\bm{s})=2^{-k_m}\prod_{j=1}^N (1-c_{mj}s_j)\;,\;\;m\in\{1,\ldots,M\}\;. \label{Kms}
\end{equation}
We have $K_m \in [0,1]$ for all $\bm{s}\in [-1,1]^N$. It is easy to check that 
$K_m = 0$ only for those $s_i \in \{-1,+1\}$ values for which the corresponding  
$x_i$-s satisfy clause $C_m$ (otherwise we always have $K_m > 0)$ . 
That is, $K_m$ plays the role of an energy function for
clause $C_m$ and its ground state value of $K_m=0$ is reached if 
$C_m$ is TRUE, and only then.
We also need the quantities 
$K_{mi} = K_m/(1-c_{mi}s_i)$ that is, with the $i$-th term missing from the product in
(\ref{Kms}). Clearly, $K_{mi} \in [0,1/2]$. The continuous time dynamical system introduced
in \cite{NatPhys_ET11} is defined via the set of $(N+M)$ ordinary differential 
equations (ODEs):
\begin{eqnarray}
&&\frac{ds_i}{dt} = \sum_{m=1}^{M} 
2 a_m c_{mi} K_{mi}(\bm{s}) K_m(\bm{s}),\;\;i=1,\ldots,N \quad \label{sdyn}  \\
&& \frac{da_m}{dt} =  a_m K_m(\bm{s}),\;\;\;\;m=1,\ldots,M\;, \label{adyn}
\end{eqnarray}
with the only requirements that $s_i(0) \in [-1,1]$, $\forall i$ and
 $a_m(0) > 0$, $\forall m$. The latter implies from
(\ref{adyn}) that $a_m(t) > 0 $, $\forall m,t$.
It was shown in Ref \cite{NatPhys_ET11} that system (\ref{sdyn}-\ref{adyn}) always finds the
solutions to $k$-SAT problems (encoded via the $\bm{C}$ matrix), when they exist, from
almost all initial conditions (the exception being a set of Lebesgue measure zero). Here
we give an intuitive picture for why that is the case. 
Due to (\ref{adyn})  the auxiliary variables $a_m$ grow exponentially at rate 
$K_m$. That is, the further is $K_m$ from its ground-state value of 0, the faster $a_m$ grows
(in that instant). Moreover, the longer has $K_m$ been away from zero, the larger is $a_m$,
as seen from the formal solution to (\ref{adyn}): $a_m(t) = 
a_m(0)\exp \left(\int_{0}^t d\tau K_m\right)$.
Equation (\ref{sdyn}) can equivalently be written as a gradient descent on an energy landscape
$V(\bm{s},\bm{a})$, that is $d \bm{s}/dt = - \nabla_s V$, where $\nabla_s$ is the gradient 
operator in the spin variables and $V(\bm{s},\bm{a}) = \sum_m a_m K_m^2(\bm{s})$\;. 
Clearly, $V \geq 0$ $\forall t$ and $V=0$ if and only if $\bm{s}$ is a $k$-SAT solution, i.e.,
satisfies all the clauses ($K_m(\bm{s}) = 0$, $\forall m$). 
\begin{figure*}[htbp] \begin{center}
\includegraphics[width=0.9\textwidth]{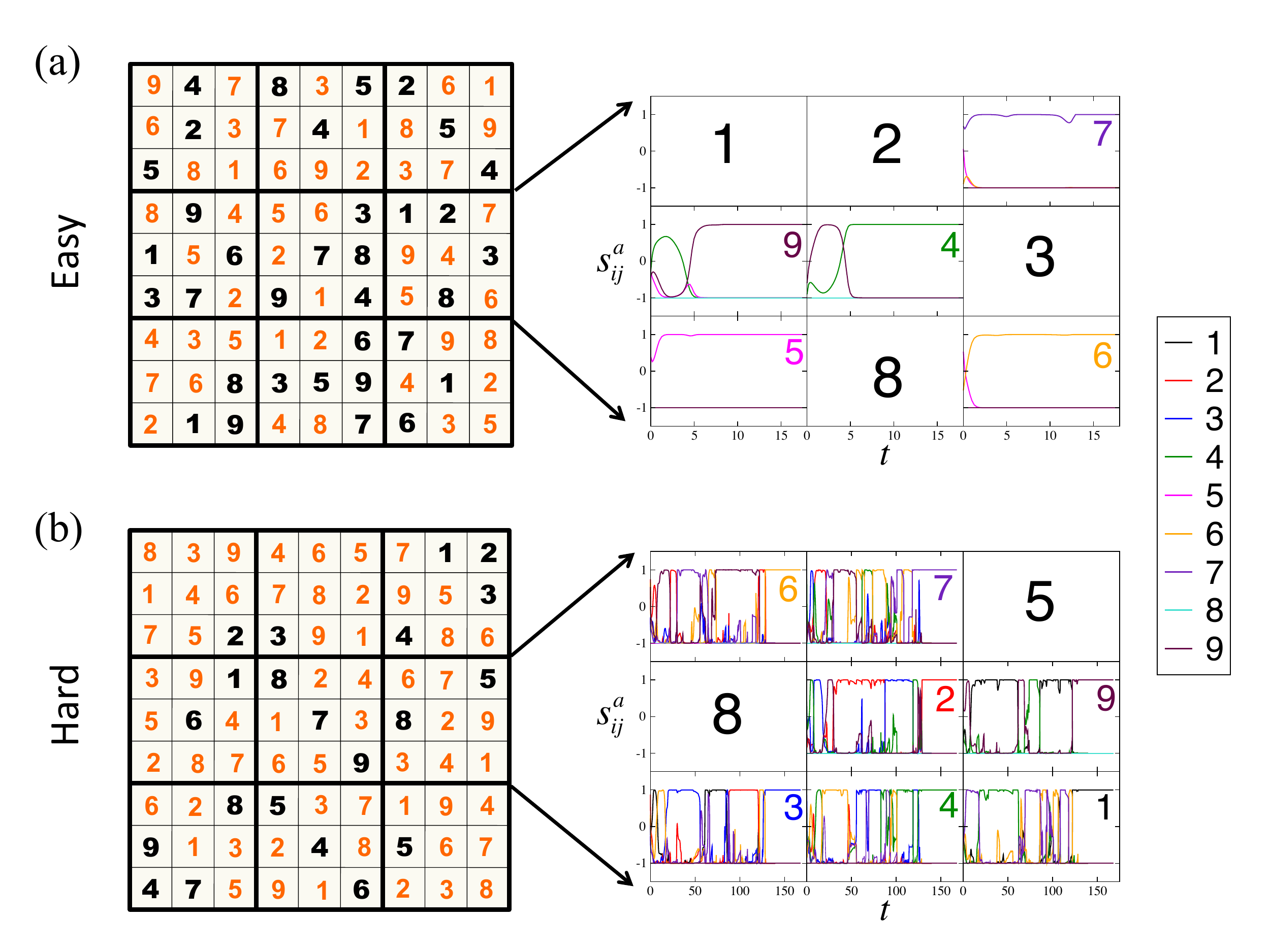}
\caption{{\bf Solving Sudoku puzzles with the deterministic continuous-time solver
(\ref{sdyn}-\ref{adyn})}. (a) presents an easy puzzle with the evolution 
of the continuous-time dynamics 
shown within a $3\times3$ grid (rows 4-6, columns 7-9). (b) shows the
same, but for a known, very hard puzzle called {\em Platinum Blonde} 
\cite{Wiki_karadimov}. } \label{fig2} 
\vspace*{-0.5cm} \end{center} \end{figure*}
From the behavior of the $a_m$ 
variables  discussed above it also follows that the least satisfied constraints will dominate $V$ 
(terms with the largest $a_m$-s). Without restricting generality, let the $a_1K_1^2$ term 
be the most dominant at  $t$. Then keeping only the dominant term on the rhs of (\ref{sdyn})
for those $i$ for which $c_{1i} \neq 0$ we get $ds_i/dt = 2a_1 c_{1i}(1-c_{1i}s_i)(K_{1i})^2$ 
or, equivalently: $d(1-c_{1i}s_i)/dt = - (1-c_{1i}s_i) 2 a_1(K_{1i})^2 $. This shows that the term
$(1-c_{1i}s_i)$ is driven exponentially fast towards zero, that is towards satisfying $K_1$ 
(and all the other constraints containing this term). As $K_1$ decreases, some other constraint
becomes dominant, and thus, in a continuous fashion, all constraints are driven towards 
satisfiability. The exponential growth guarantees that the trajectory is always pulled out of any
potential well. When the problem is unsatisfiable, the system generates a  
chaotic dynamics in $[-1,1]^N$, indefinitely. For more details about the properties of the
CTDS (\ref{sdyn}-\ref{adyn})  see Ref. \cite{NatPhys_ET11}.

\subsection*{Puzzle hardness as transient chaotic dynamics}

Since Sudoku puzzles always have a solution, the corresponding Boolean 
SAT CNF formulation also has a solution, and system (\ref{sdyn}-\ref{adyn}) 
will always find it. The nature of the dynamics, however will depend on the 
hardness of the puzzle as we describe next.

In Fig.\ref{fig2}a we show an easy puzzle with $34$ clues (black numbers)
\cite{SudokuOftheDay}. 
After transforming this problem into SAT, we obtain $N=126$ and $M=717$, 
with a constraint density of $\alpha = M/N=5.69$. 
As described above, in our implementation there is a
spin variable $s_{ij}^a$ associated to every Boolean variable $x_{ij}^a$ in every 
3D cell $(i,j,a)$. 
In the right panels of Fig. \ref{fig2} we show the dynamics of the spin 
variables in the cells of the $3\times 3$ grid formed by rows 4-6 and  columns 7-9. 
The $s_{ij}^a(t)$ curves are colored by the digit $a$ they represent ($a=1,\ldots,9$) 
as indicated in the color legend of Fig \ref{fig2}. The dynamics was started 
from a random initial condition. Indeed, our solver finds the solution very quickly, 
for the easy puzzle in Fig.\ref{fig2}a.
\begin{figure*}[htbp] \begin{center}
\includegraphics[width=0.95\textwidth]{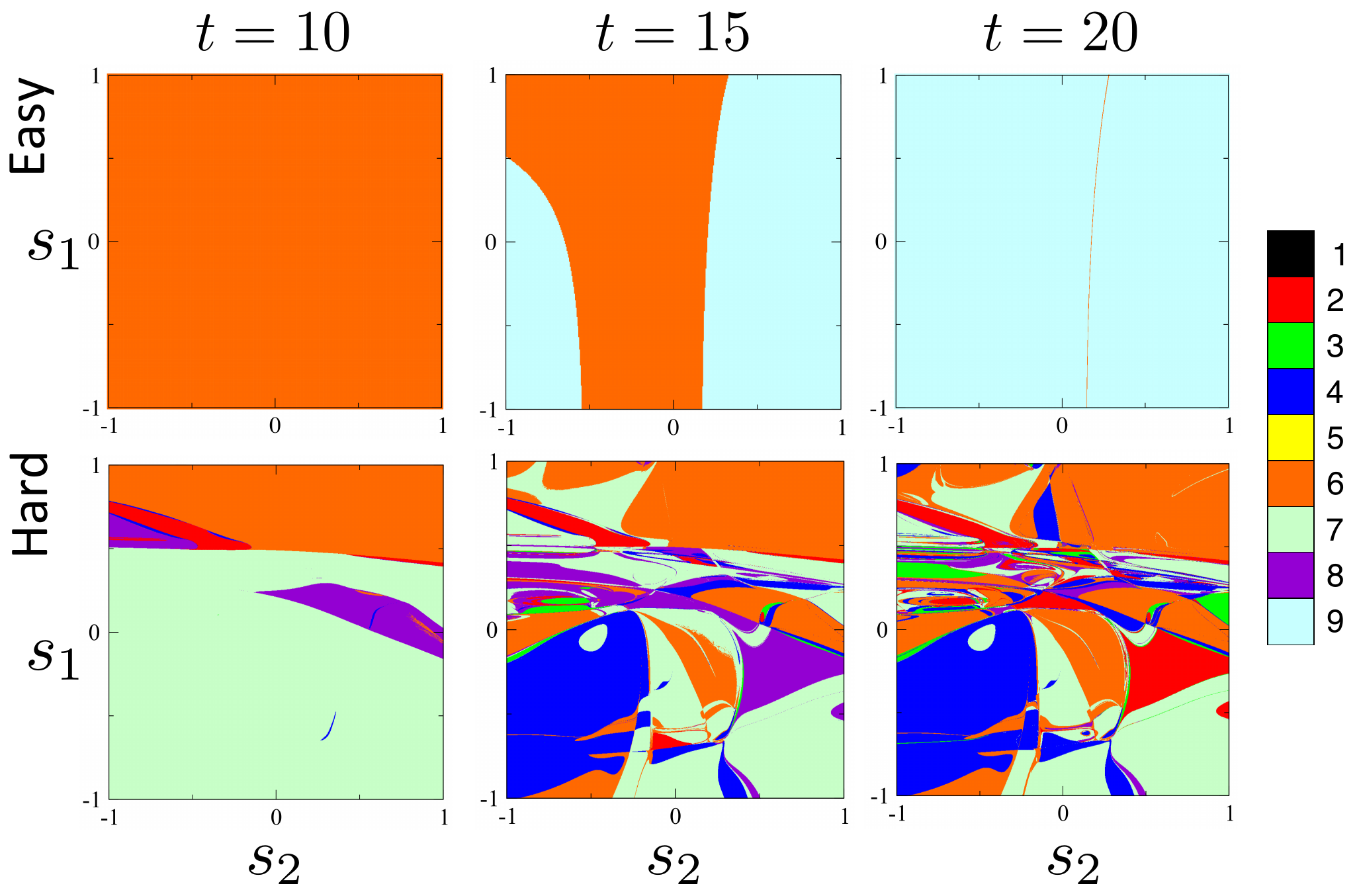}
\caption{{\bf Puzzle hardness as chaotic dynamics.} We color
the points of a $10^3\times 10^3$ grid in an arbitrary plane $(s_1,s_2)$ 
at time instant $t$ according to the digit $D_{pq}$ the solver is considering 
in an arbitrary but fixed cell $(p,q)$ at that instant, given that we started 
the trajectory of the CTDS from
those grid-points. For these initial conditions only the points in the 
$(s_1,s_2)$ plane were varied, all other spin values were kept fixed at the same
randomly chosen values. For an easy problem (top row of panels), and for $(p,q) = (1,1)$ 
almost all
initial conditions in this plane involve only two digits, and after $t=20$ the corresponding
trajectories have converged to the solution digit (9, light blue), except for a thin line,
which, however, will also become light blue. The bottom row
of panels shows the same for a hard problem based on what happens in the cell 
$(p,q) = (6,8)$. The strong
sensitivity to initial conditions appears as fractal structures of increasing complexity
as time goes on, before eventually everything converges to the same color/digit 
(not shown).} \label{fig3} 
\vspace*{-0.5cm} \end{center} \end{figure*}

In Fig. \ref{fig2}b  we show the dynamical evolution of variables for a very hard 
Sudoku instance with only $21$ clues. This puzzle  has been listed as one of the 
world's hardest Sudokus, and even has a special name: ``Platinum Blonde" 
\cite{PlatinumBlonde,Wiki_karadimov}, and it 
was the most ``difficult" for our solver among all the puzzles we tried. 
After transforming it into SAT CNF, we obtain $N=257$ variables and $M=2085$ 
constraints. 
Not only that we have twice as many unknown variables but the constraint density 
$\alpha = M/N=8.11$ is also larger than in the previous case, 
signaling the hardness of the corresponding SAT instance. 
The complexity of the dynamics in this case is seen in the right panel of 
Fig. \ref{fig2}b, exhibiting long chaotic transients before the solution is found 
at around $t\simeq 150$. For an animation of the dynamics for a similarly hard 
puzzle \cite{PNAS_ERT07} see Ref \cite{Anim}. 

We can also observe from the right panels in Fig \ref{fig2} that there is 
one dominating digit ($a$-value), corresponding to which vertical cell at 
that given $(i,j)$ grid cell 
has the largest 
$|s_{ij}^{a}|$ value. This 
 can be taken as the digit $D_{ij}$ the 
solver is considering in the given  grid cell  $(i,j)$ at that moment.  We will use this observation
to provide below an alternate illustration of the dynamics' transiently chaotic behavior.
Let us fix a random initial condition except for two chosen variables 
that are varied along the points of a square grid within the domain $[-1,1]^2$. 
There is no particular relevance as to which pairs of variables are chosen to be varied, 
let us denote them by $s_1$ and $s_2$. Let us choose an arbitrary empty cell $(p,q)$ 
in the original Sudoku puzzle  and monitor the dominating digit in it at time $t$. We will
color the initial conditions in the plane $(s_1,s_2)$ according to the dominating
digit in $(p,q)$ at time $t$. This will provide a map expressing the 
``sensitivity to initial conditions'' that varies across time. Since all puzzles have 
solutions, the maps eventually assume one solid color according to the digit of 
the solution in the monitored cell, however, 
for hard puzzles, it may assume highly complex patterns before it does that,
as shown in Fig.  \ref{fig3}. 
In Fig.\ref{fig3}  we show these colormaps for the easy and hard Sudoku 
 puzzles shown in Fig.\ref{fig2} at  times $t=10, 15, 20$. For the easy 
 puzzle (top row of panels) 
 the cell was chosen to be $(p,q) = (1,1)$. At time $t=10$ the whole map shows 
 $D_{1,1}=6$ (orange), which is not the solution digit (it is still searching for the solution). 
 At time $t=15$, however,  we see two clearly separated domains, in one 
 of them $D_{11}=6$, in the other  $D_{11}=9$ (cyan) and the latter is the 
 correct digit. As time passes, the orange (incorrect) domain shrinks, because 
trajectories from an increasing number of initial conditions find the solution.  
 At $t=20$  almost the whole map shows the correct digit $D_{11}=9$, except for 
 a thin line. 

In the case of the hard Sudoku puzzle (bottom row in Fig.  \ref{fig3}, $(p,q) = (6,8)$) 
more colors enter the picture with time, in a 
complex fractal-like pattern. On this fractal set changing the initial condition slightly
may result in a completely different digit (color) being considered in cell $(p,q)$ at time
$t$. 
This sensitivity to initial conditions is indicative of the chaotic behavior of the 
(deterministic) search dynamics.

The appearance of transient chaos is a fundamental feature of the search dynamics
and can be used to separate problems by their hardness. 
In Ref \cite{NatPhys_ET11} we have shown that
within the {\em thermodynamic limit} ($N \to \infty$, $M\to \infty$, $\alpha = M/N = \mbox{const.}$) 
of random $k$-SAT ensembles this appears as a phase transition at the so-called chaotic transition point $\alpha_{\chi}$ in terms of the constraint density $\alpha=M/N$. Since there is
no ``thermodynamic limit'' for $9\times 9$ Sudoku problems ($N < 729$), one cannot 
define a simple order-parameter and use it to rate problem hardness in the same way \cite{NatPhys_ET11}. However, once a problem is given, the corresponding
dynamical system (\ref{sdyn}-\ref{adyn}) is well defined, and so is its dynamical 
behavior. Even though we do not have a well-defined ensemble-based 
statistical order parameter,
(which has little meaning for specific SAT instances anyway), here we show next how can we 
use a well-known invariant quantity from non-linear dynamical system's theory to categorize problem 
hardness for specific instances.

\subsection*{A Richter-type scale for Sudoku hardness}

\begin{figure*}[htbp] \begin{center}
\includegraphics[width=0.80\textwidth]{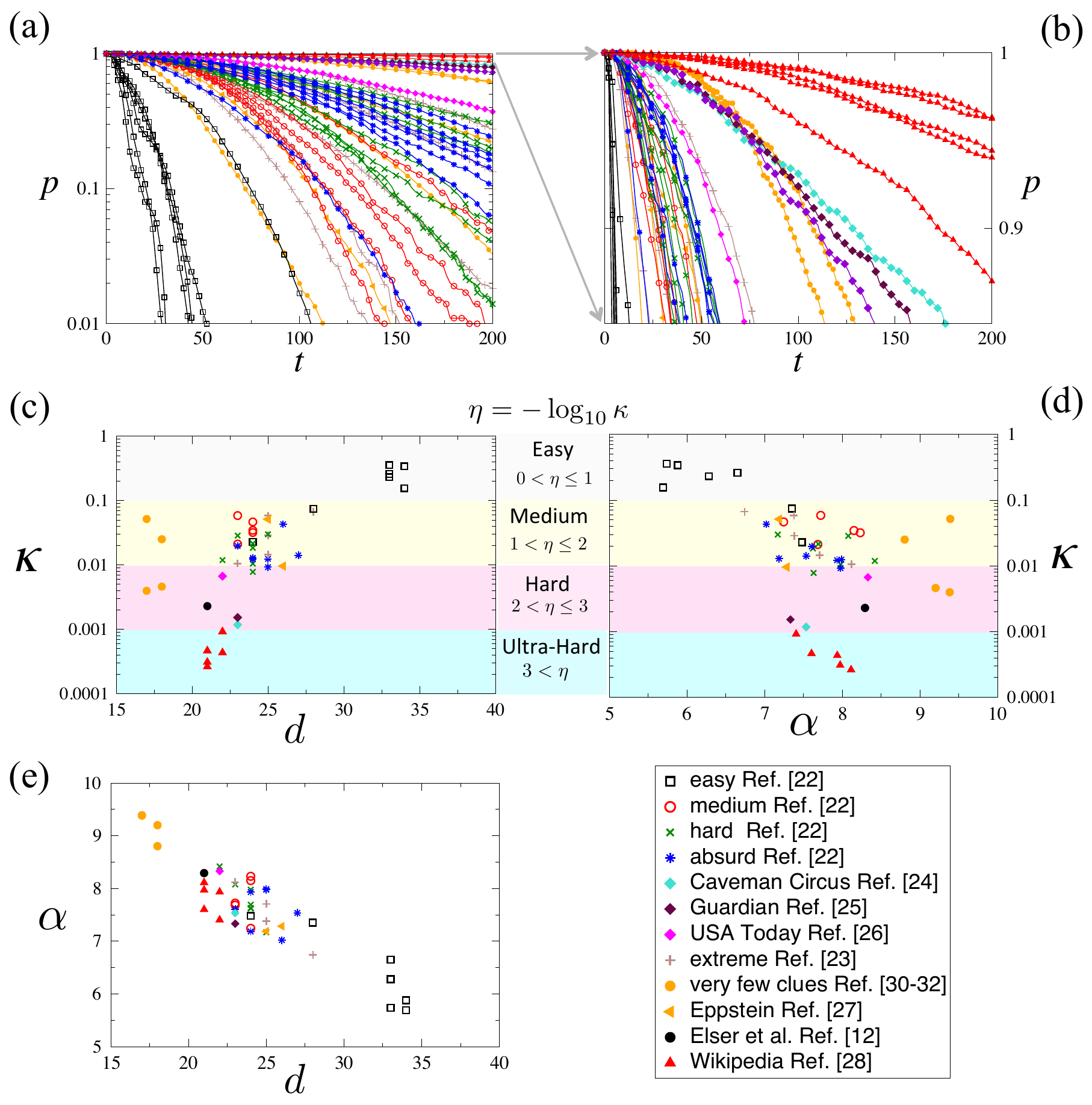}
\caption{{\bf Escape rate as hardness indicator.} (a) shows the distribution in log-linear
scale of the fraction $p(t)$ of $10^4$ randomly started
trajectories of (\ref{sdyn}-\ref{adyn})  that have not yet found a solution by analog time $t$ for a 
number of Sudoku puzzles taken from the literature (see legend and text) 
with a wide range of human difficulty ratings. The escape rate is obtained
from the best fit to the tail of the distributions. (b) is a magnification of (a)
for hard puzzles. (c) and (d) show the escape rate $\kappa$ in semilog 
scale vs the number of clues 
$d$ and constraint density $\alpha$ indicating good correlations with 
human ratings (color bands).
(e) shows the relationship between the number of clues $d$ and 
$\alpha$ for the puzzles considered.
} \label{fig4} 
\vspace*{-0.5cm} \end{center} \end{figure*}
As suggested by the two examples in Fig \ref{fig3}, the hardness of Sudoku puzzles 
correlates with the length of chaotic transients. A consistent way to characterize 
these chaotic transients is to plot the distribution of their lifetime. Starting trajectories 
from many random initial conditions, let $p(t)$ indicate the probability that the dynamics 
has not found the solution by analog time $t$.  A characteristic property of transient 
chaos \cite{TransientChaosBook,ChaosBook} in hyperbolic dynamical systems is that $p(t)$ 
shows an exponential decay: $p(t)\sim e^{-\kappa t}$, where $\kappa$ is 
called the escape rate. The escape rate, an easily measurable quantity, theoretically
can be expressed as a zero of the spectral determinant
of the evolution operator corresponding to the dynamical system (\ref{sdyn}-\ref{adyn})
and well approximated using the machinery of cycle expansions based on dynamical 
zeta functions \cite{ChaosBook}. It is an invariant measure of the dynamics in the sense
that it characterizes solely the chaotic non-attracting set in the phase space of the system,
and  it
does not depend on the distribution of the initial conditions, its support, or 
the details of the region from where the escape is measured (as long as it contains the
non-attracting set) \cite{TransientChaosBook}. 

In Fig. \ref{fig4}a we plot the distribution $p(t)$ in log-linear scale  
for several puzzles gathered form the literature. The distributions were obtained from over $10^4$
random initial conditions. The decay shows a wide range of variation between the puzzles.
For easy puzzles the transients are very short, $p(t)$ decays  fast resulting in  large 
escape rates but for hard puzzles $\kappa$ can be very small.  
Fig. \ref{fig4}b shows a zoom onto the $p(t)$ of  hard puzzles. 
In spite of the large variability of the decay rates, we see 
that in all cases the escape is exponentially fast or faster (the curves in Figures  
\ref{fig4}a,b are straight lines or bend downward).

The  several orders of magnitude variability of $\kappa$  naturally behooves us to
use a logarithmic measure of $\kappa$ for puzzle hardness, see Fig.\ref{fig4}c, which
shows the escape rates on a semilog scale as function of the number of clues, $d$. 
Thus, the escape rate can be used to define a kind of ``Richter''-type scale for Sudoku hardness: 
\begin{equation}
\eta=-\log_{10}(\kappa) \label{richter}
\end{equation}
with easy puzzles  falling in the range $0< \eta \leq 1$, medium ones in $1<\eta\leq 2$, 
hard ones in $2<\eta\leq 3$ and for ultra-hard puzzles $\eta > 3$.

We chose several instances from the ``Sudoku of the Day'' website \cite{SudokuOftheDay} in 
four of the categories defined there: easy (black square), medium (red circle), hard (green x) 
and absurd (blue star). These ratings on the website try to estimate the 
hardness of puzzles when solved by humans. These ratings correlate 
very well with our hardness measure $\eta$, giving an  
average hardness value of $\langle \eta \rangle = 0.816$ for easy, 
$\langle \eta \rangle =1.439$ medium, $\langle \eta \rangle = 1.782$ for hard and 
$\langle \eta \rangle =1.809$ for what they call absurd. 
Another site we analyzed puzzles from is
``Extreme Sudoku'' \cite{ExtremeSudoku} (brown $+$ signs on Fig.\ref{fig4}).  
It claims to offer extremely hard Sudoku puzzles, their categories being:  
evil, excessive, egregious, excruciating and extreme. Indeed those puzzles 
are difficult with a range of $\eta \in [1.1,1.9]$ on the hardness scale, however, 
still far from the hardest puzzles we have found in the 
literature. Occasionally,  daily newspapers  present  puzzles claimed to be the 
hardest Sudoku puzzles of the year. In particular, the escape rate for the 
Caveman Circus 2009 winner \cite{Caveman} (turquoise diamond) and 
the Guardian 2010 hardest puzzle \cite{Guardian} (maroon diamond) 
are indeed one order of magnitude smaller than the hardest puzzles 
on the daily Sudoku websites, placing them at $\eta=2.93$ and $\eta=2.82$ 
on the hardness scale. The USA Today 2006 hardest puzzle \cite{USAtoday}, 
however, does not seem to be that hard for our algorithm having $\eta=2.17$ (magenta diamond). 
Eppstein \cite{Eppstein} gives two Sudoku examples (orange left-pointing triangles) while describing his algorithm, one with
$\eta=1.288$ and a much harder one with $\eta=2.017$. Elser {\it et al.}   
 \cite{PNAS_ERT07} present an extremely hard Sudoku (black filled circle), 
 which has an escape rate of $\kappa=0.0023$ resulting in $\eta=2.639$.
 
 The smallest escape rates we have found are for the  Sudokus listed 
as the hardest on Wikipedia \cite{Wiki_karadimov,Wikihardest} (red triangles). 
The five puzzles, which we tested 
are called Platinum Blonde, Golden Nugget, Red Dwarf, coly013 and tarx0134. 
They have a hardness in the range $3<\eta<3.6$, 
the Platinum Blonde (shown in Fig.\ref{fig2}b) being the hardest with $\eta=3.5789$ 
(corresponding to an escape rate of $\kappa=0.00026$).

While the escape rate correlates surprisingly well with human ratings of Sudoku hardness, 
it is natural to expect a correlation with the number of clues, $d$. Indeed, as a general 
rule of thumb, the fewer clues are given, the harder the puzzle, however, this is not 
universally true \cite{RT11}. 
Here we tested a few instances with  minimal \cite{arxiv_MTC12}, that is $17$ clues
and almost minimal $18$ clues (orange filled circles) 
\cite{VeryFewClues1,VeryFewClues2,VeryFewClues3}. As seen from Fig.\ref{fig4}c, 
these are actually easier ($1.2<\eta<2.4$) than the hardest instances with more $d=21,22$ clues.  
In Fig.\ref{fig4}d we then plot the escape rate  as 
function of the constraint density $\alpha = M/N$, leading to practically the same conclusion. 
This is because  the constraint density $\alpha$ is essentially linearly correlated with the 
number of givens $d$, as shown in Fig.\ref{fig4}e. 
The apparent non-monotonic behavior of puzzle hardness with the number of givens, 
(or constraint density) is due to the fact that hardness cannot simply be characterized by
a global, static variable such as $d$ or $\alpha$, but it also depends on the {\em positioning}
pattern of the clues, as also shown by concrete examples in Ref \cite{RT11}.

\section*{\large Discussion}

Using the world of Sudoku puzzles, here we have presented further evidence that 
optimization hardness translates into complex dynamical behavior by an algorithm 
searching for solutions in an optimal fashion. Namely, there seems to be a trade-off 
between algorithmic performance and the complexity of the algorithm and/or its behavior.
Simple, sequential search algorithms have a trivial description and simple dynamics,
but an abysmal worst-case performance ($2^N$), whereas algorithms that
are among the best performers are complex in their description (instruction-list) 
and/or behavior (dynamics). This happens because in order to improve performance, 
algorithms have to exploit the structure of the problem one way or another. As hard
problems have complex structures, the dynamics of the algorithms should be indicative
of the problem's hardness. However, as a word of caution, observing
complex dynamics performed by some black-box algorithm does not necessarily 
imply problem hardness. For example, one could consider any arbitrary, but ergodic 
dynamical system with complex behavior 
in the same state space as the problem's. Ergodicity guarantees the algorithm 
to eventually visit all of the $2^N$ states, 
and hence to always find solutions.  
But its instruction list would have no relevance to the problem itself (apart from the
checking instructions to see if the new state satisfies the problem) and thus, 
it could take long times to find solutions even for problems that are otherwise 
easily solved by other algorithms. Hence, dynamical properties can only be regarded as
descriptors of problem hardness if they are generated by algorithms that: 1) exploit the
structure of the state space of the problem and 2) they show similar or better performance 
compared to other algorithms on the same problems. 

The continuous-time dynamical system \cite{NatPhys_ET11}  (\ref{sdyn}-\ref{adyn}) as
a {\em deterministic algorithm} does have these features: 1) the search happens 
on an energy landscape $V=\sum_m a_m K_m^2$ that incorporates simultaneously 
{\em all} the constraints (problem structure) 2) it solves easy problems 
efficiently (polynomial time, both analog and discrete) and 3) it guarantees 
to find solutions to hard problems even for solvable cases where many other algorithms 
fail. Although it is not a polynomial cost algorithm, it seems to find solutions in 
continuous-time  $t$ that scales polynomially with $N$ \cite{NatPhys_ET11}. These
features and the fact that the algorithm is formulated as a deterministic 
dynamical system with continuous variables, allows us to apply the theory of 
nonlinear dynamical systems on CTDS (\ref{sdyn}-\ref{adyn}) 
to characterize the hardness of Boolean satisfiability problems. In particular, via the
measurable escape rate $\kappa$, or its negative log-value $\eta$, 
we can provide a single-scalar measure of hardness, well defined for any {\em finite 
instance}. We have illustrated this here on Sudoku puzzles, but the analysis can be
repeated on any other ensemble from NP. Having 
a mathematically well-defined number to characterize 
optimization hardness for specific problems in NP  provides more information 
than the polynomial/exponential-time solvability classification, or knowing what the 
constraint density $\alpha = M/N$ is (the latter being a non-dynamic/static measure).
Moreover, within the framework of CTDS (\ref{sdyn}-\ref{adyn}), 
dynamical systems and chaos theory methods can now be brought forth to help 
develop a novel understanding of optimization hardness.

\section*{\large Methods}

Here we continue to describe in detail how a Sudoku puzzle is transformed into a SAT 
problem in CNF. 

Type 1) constraints (main text) impose the uniqueness of the symbol 
$D_{ij}$ in a given cell, expressed as a $+1$-in-9-SAT constraint:   
\begin{equation}
(x_{ij}^1,x_{ij}^2,\dots,x_{ij}^9)\;. \label{uniqueS}
\end{equation}
Having $9\times 9$ cells in the puzzle, this gives in total $81$, $+1$-in-9-SAT constraints. 

Type 2) constraints on rows, columns and sub-grids further impose that in every 
layer $L_a$ we have the following $27$,  $+1$-in-9-SAT constraints:
\begin{eqnarray}
&&\mbox{Rows:} \nonumber \\
&&\qquad(x_{i1}^a, x_{i2}^a,\dots, x_{i9}^a),   \,\, i=1,\dots,9  \\
&&\mbox{Columns:} \nonumber \\
&&\qquad(x_{1j}^a, x_{2j}^a,\dots, x_{9j}^a)  \,\, j=1,\dots,9 \\
&&\mbox{Subgrids:} \nonumber \\
&&\qquad(x^a_{m+1,n+1}, x^a_{m+1,n+2},x^a_{m+1,n+3}, \nonumber \\
&&\qquad x^a_{m+2,n+1}, x^a_{m+2,n+2},x^a_{m+2,n+3}, \\
&&\qquad x^a_{m+3,n+1}, x^a_{m+3,n+2},x^a_{m+3,n+3})\;. \;\;\;m,n=0,3,6 \nonumber
\end{eqnarray}
Together with the $81$ constraints of type 1) we thus have in total
 $9\times 27+81=324$  constraints in +1-in-9-SAT form.  

Finally, type 3) constraints are imposed via $d$ given digits or clues. It was only recently shown
that uniqueness of a solution demands that $d \geq 17$  \cite{arxiv_MTC12}.
As discussed in the main text, each clue will eliminate  $4$ constraints: 
in its vertical tower, its column,
its row and the $3\times 3$ sub-grid containing the clue.
For example, let us examine layer $L_4$ (Fig.\ref{fig1}c) of the puzzle shown 
in Fig.\ref{fig1}a. There are three clues of $4$ in cells $(1,9)$, $(3,3)$, 
$(4,4)$ and thus $x_{1,9}^4=1$, $x_{3,3}^4=1$, $x_{4,4}^4=1$ have to be
fixed as TRUE in $L_4$. In order to satisfy the constraints, the other variables in the 
same rows, columns, blocks and  vertical columns must be set to FALSE.
The unknown variables left in the SAT problem will be those in the light cells of Fig.\ref{fig1}c. 
(The other clues will eliminate constraints and variables in other layers and 
vertical columns.) The total number of unknown variables $N$ depends on 
$d$ and on the placement of clues. The number of constraints is always 
$324-4d$, however the number of variables in a clause can vary. For 
example in Fig.\ref{fig1}c the constraint corresponding to the second row 
in layer $L_4$ has only $2$ unknown variables left ($+1$-in-$2$-SAT). 

After the unknown Boolean variables and the  constraints  have been identified 
we need to transform the formula into CNF. There are several ways of 
doing this, here we use the following general procedure. A $+1$-in-$k$-SAT 
clause defined on the $(y_1, y_2,\dots,y_k)$ variables  can be written  as one  
$k$-SAT and $k(k-1)/2$  of $2$-SAT constraints:
\begin{equation}
(y_1 \vee y_2 \vee \dots \vee y_k) \wedge
\left[ \bigwedge_{i<j}(\overline{y}_i \vee \overline{y}_j )  \right]
\end{equation} 
The disjunction ($\vee$) of the first $k$ variables enforces that at least one variable 
must be true, but the rest of ($k(k-1)/2$) $2$-SAT type constraints 
ensure that only one of them is allowed to be true.

\section*{\large Acknowledgments}
This work was supported in part by a grant of the Romanian  
National Authority for Scientific Research, CNCS-UEFISCDI,  grant number
PN-II-RU-TE-2011-3-0121 (MER) and by a University of Notre Dame internal capitalization 
grant (ZT).


\begin{thebibliography}{99}


\bibitem{RT11} Rosenhouse, J. \& Taalman, L. {\em Taking Sudoku Seriously: The Math Behind
the World's most Popular Pencil Puzzle} (Oxford University Press,  New York, 2011). 

\bibitem{GJ90} Garey, M. R. \& Johnson, D. S. {\em Computers and Intractability: 
A Guide to the Theory of NP- Completeness} (W. H. Freeman \& Co., New York, NY, USA, 1990).

\bibitem{K72} Karp, R.M.  Reducibility among combinatorial problems. 
In {\em Complexity of Computer Computations.}, R.E. Miller and J.W. Thatcher (editors).  
{\em Proc. of a Symp. on the Complexity of Computer Computations.} (New York: Plenum. pp. 
85-103, 1972).

\bibitem{IEICE_YS03} Yato, T. \& Seta, T. Complexity and completeness of 
finding another solution and its
application to puzzles. {\em IEICE Trans. Fundamentals} E86-A(5), 1052-1060 (2003).

\bibitem{CommACM_F09} Fortnow, L. The status of the P versus NP problem. 
{\em Commun. ACM} {\bf 52}, 78-86 (2009).

\bibitem{JPhysA_B82} Barahona, F. On the computational complexity of Ising spin glass models.
{\em J. Phys. A: Math. Gen.} {\bf 15}, 3241-3253 (1982).

\bibitem{STOC_I00} Istrail, S. Statistical Mechanics, Three-Dimensionality and NP-Completeness: I. Universality of Intractability of the Partition Functions of the Ising Model Across Non-Planar Lattices. 
{\em Proceedings of the 32nd ACM Symposium on the Theory of Computing (STOC00)}, ACM Press, 
pp. 87-96 (2000)

\bibitem{TheorCompSci_BK04} Brueggemann, T. \& Kern, W. 
An improved local search algorithm for 3-SAT.
{\em Theor. Comp. Sci.} {\bf 329}(1-3), 303Ð313 (2004).

\bibitem{OttBook} Ott, E. {\em Chaos in Dynamical Systems} 2nd edn (Cambridge Univ. Press, 2002).

\bibitem{ChaosPhys} Cencini, M., Cecconi, F. \& Vulpiani, A. {\em Chaos: from 
simple models to complex systems} (World Scientific, Singapore, 2009).

\bibitem{TransientChaosBook}  Lai, Y.-C. \& T\'el, T. 
{\em Transient Chaos: Complex Dynamics on Finite-Time Scales}
(Springer 2011).

\bibitem{PNAS_ERT07} Elser, V., Rankenburg, I. \& Thibault, P. 
Searching with iterated maps.
{\em Proc. Natl. Acad. Sci. USA}  {\bf 104}, 418-423 (2007).

\bibitem{NatPhys_ET11} Ercsey-Ravasz, M, \& Toroczkai, Z. 
Optimization hardness as transient chaos in an analog approach to constraint satisfaction.
{\em Nature Physics} {\bf 7}, 966-970 (2011). 

\bibitem{PNAS_KT84} Kadanoff, L. P. \& Tang, C. Escape from strange repellers. 
{\em Proc. Natl Acad. Sci.} {\bf 81}, 1276-1279 (1984).

\bibitem{PhysRep_TL08} T\'el, T. \& Lai, Y-C. Chaotic transients in spatially extended systems. {\em 
Phys. Rep.} {\bf 460}, 245-275 (2008).


\bibitem{PRL_ZM08}
Zdeborov\'a, L. \& M\'ezard, M. Locked constraint satisfaction problems. {\em Phys. Rev. Lett. } {\bf 101}, 
078702 (2008).

\bibitem{JSMTE_ZM08}
Zdeborov\'a, L. \& M\'ezard, M. Constraint satisfaction problems with
isolated solutions are hard. {\em J. Stat. Mech.: Theor. Exp.} P12004 (2008).

\bibitem{PlatinumBlonde} \url{http://forum.enjoysudoku.com/the-hardest-sudokus-new-thread-t6539.html}

\bibitem{Wiki_karadimov}  \url{http://wiki.karadimov.info/index.php/Sudoku_algorithms#Exceptionally_difficult_Sudokus_.28hardest_Sudokus.29}

\bibitem{Anim} \url{http://www.youtube.com/watch?v=y4_aSLP9g_w}


\bibitem{ChaosBook} Cvitanovi\'{c}, P., Artuso, R., Mainieri, R. Tanner, G. \& Vattay, G. {\em 
Chaos: Classical and
Quantum}, ChaosBook.org/version13 (Niels Bohr Institute, Copenhagen 2010).

\bibitem{SudokuOftheDay} \url{http://www.sudokuoftheday.co.uk}


\bibitem{ExtremeSudoku} \url{http://www.extremesudoku.info/sudoku.html}

\bibitem{Caveman} \url{http://cavemancircus.com/2009/11/05/the-hardest-sudoku-puzzle-ever/}

\bibitem{Guardian} \url{http://www.guardian.co.uk/media/2010/aug/22/worlds-hardest-sudoku}

\bibitem{USAtoday} \url{http://www.usatoday.com/news/offbeat/2006-11-06-sudoku_x.htm}

\bibitem{Eppstein} Eppstein, D. Solving Single-digit Sudoku Subproblems, \url{http://arxiv.org/abs/1202.5074v2} 

\bibitem{Wikihardest} \url{http://en.wikipedia.org/wiki/Algorithmics_of_sudoku#Exceptionally_difficult_Sudokus_.28hardest_Sudokus.29}

\bibitem{arxiv_MTC12} McGuire, G., Tugeman, B. \& Civario, G. 
There is no 16-Clue Sudoku: Solving the Sudoku
Minimum Number of Clues Problem. \url{http://arxiv.org/abs/1201.0749}


\bibitem{VeryFewClues1} \url{http://mapleta.maths.uwa.edu.au/~gordon/sudokumin.php}

\bibitem{VeryFewClues2} \url{http://en.wikipedia.org/wiki/File:Symmetrical_18_clue_sudoku_01.JPG}

\bibitem{VeryFewClues3} \url{http://www.flickr.com/photos/npcomplete/3603730706/}



\end{thebibliography}
\end{document}